# A Study of Transition of the Expansion of the Universe from a Phase of Deceleration to Acceleration through a Conversion of Matter into Dark Energy in the Framework of Brans-Dicke Theory

Sudipto Roy

Department of Physics, St. Xavier's College, 30 Mother Teresa Sarani (Park Street),

Kolkata 700016, West Bengal, India.

roy.sudipto1@gmail.com

## Abstract

The present study is based on a generalized form of Brans-Dicke (BD) theory where, the dimensionless BD parameter is regarded as a function of the scalar field, which is reciprocal of the gravitational constant. The field equations have been solved by incorporating an empirical function f(t) in the expression representing the conservation of matter. This function f(t) has been chosen to account for a conversion of matter (both dark and baryonic) into some other form, possibly dark energy, which is known to be responsible for the accelerated expansion of universe. The requirement of a signature flip of the deceleration parameter (q), which is evident from other studies, sets the boundary conditions to be satisfied by the function f(t), leading to the formulation of its time dependence. A simple empirical relation was initially assumed to represent the time dependence of f(t), and the constants in this expression have been determined from these boundary conditions. The BD parameter has been found to have a negative value throughout the range of study. The dependence of BD parameter upon the scalar field has been depicted graphically. A smooth transition of the universe, from a decelerated to an accelerated phase of expansion, is found to occur due to a conversion of matter into dark energy. The gravitational constant is found to be increasing with time.

## Introduction

From recent studies regarding the expansion of the universe it is quite evident that the universe has undergone a smooth transition from a decelerated phase to its present accelerated phase of expansion [1,2]. This expansion of the universe was initially believed to be governed solely by gravitational attraction among celestial bodies, which is capable of causing only decelerated expansion. The observation of accelerated expansion of the universe, evident from the negative value of the experimentally determined deceleration parameter, triggered speculations about the existence of a special kind of matter or energy responsible for this acceleration. Interactions of normal matter with this new form of matter/energy are believed to make the effective pressure sufficiently negative, leading to a repulsive effect. Dark energy is the name of this new matter/energy that causes accelerated expansion. A huge amount of cerebral effort has gone into the determination of its true nature. To account for the accelerated expansion of the universe, a number of theoretical models have been proposed.

In many of these models, the cosmological constant has been chosen to represent the entity named dark energy [3]. Models regarding cold dark matter (CDM) have a serious drawback in connection to the value of cosmological constant $\Lambda$. The currently observed value of Cosmological constant $\Lambda$ for an accelerating Universe does not match with that of the value in Planck scale or Electroweak scale [4]. The problem can be rendered less acute if one tries to construct dark energy models with a time dependent cosmological parameter. But there are limitations of many such models proposed by researchers [5, 6].

The scalar field models, proposed as alternative theories to the dynamical $\Lambda$ models, are the ones in which the equation of state of dark energy changes with time. Quintessence models, among the many proposed scalar field models, are the ones endowed with a potential so that the contribution to the pressure sector, can evolve to attain an adequately large negative value, thus generating the observed cosmic acceleration [7, 8]. One main drawback of these quintessence models is that most of the quintessence potentials are chosen arbitrarily and do not have a proper theoretical justification explaining their genesis. Naturally a large number of other alternative scalar field models, for example the tachyon [9, 10], k-essence [11, 12], holographic [13, 14] dark energy models have appeared in the literature with their own virtues and shortcomings.

The dark energy and cold dark matter, in most of the scalar field models, are normally allowed to evolve independently. However, there are attempts to include an interaction amongst them so that one grows at the expense of the other [15]. Non minimal coupling of the scalar field with the dark matter sector through an interference term in the action has helped in explaining the cosmic acceleration. These fields are known as ‘Chameleon fields’ and they have been found to be useful in representing dark energy [16, 17]. In the framework of Brans-Dicke theory, non minimal coupling between the scalar field and geometry can be shown to account for the accelerated expansion of the universe. A potential function term $V(\varphi)$, which is a function of the BD scalar field $\varphi$ itself, is incorporated in a modified form of the Brans-Dicke (BD) theory. This new model can serve as a strong candidate in explaining the acceleration of the Universe [18].

In the framework of the Brans-Dicke theory of gravitation, there are a number of theoretical models that can be analyzed and compared with one another. For example, Sheykhi et al. [19] worked with the power-law entropy-corrected version of BD theory defined by a scalar field and a coupling function. In another literature Sheykhi et al. [20] considered the HDE model in BD theory to think about the BD scalar field as a possible candidate for producing cosmic acceleration without invoking auxiliary fields or exotic matter considering the logarithmic correction to the entropy. Jamil et. al. [21] studied the cosmic evolution in Brans-Dicke chameleon cosmology. Pasqua and Khomenko [22] studied the interacting logarithmic entropy-corrected HDE model in BD cosmology with IR cut-off given by the average radius of the Ricci scalar curvature.

In some models based on the BD theory one finds a quintessence scalar field that can give rise to a late time acceleration for a wide range of potentials [23]. An interaction between dark matter and the BD scalar field showed that the matter dominated era can have a transition from a decelerated to an accelerated expansion without any additional potential [24]. On the other hand BD scalar field alone can also drive the acceleration without any quintessence matter or any interaction between BD field and dark matter [25].

Several such models are found to have discrepancies in the sense that the matter dominated universe has an ever accelerating expansion according to them, in contradiction with the observations. In addition to that, one needs to consider a wide range of values of the BD parameter $\omega$ to explain different phenomena. In order to explain the recent observation of acceleration, many of the models require a very low value of the BD parameter $\omega$ of the order of unity whereas the local astronomical experiments demand a very high value of $\omega$ [26].

In the present study we have made an attempt to explain the transition from a decelerated to an accelerated phase of expansion of the universe by taking into account a possibility of a time varying matter content of the universe which can theoretically be attributed to an inter-conversion between matter (both dark and baryonic) and dark energy, since dark energy is the name of the entity causing the accelerated expansion. Our study reveals that the deceleration parameter changes sign from positive to negative when we have our total matter content (dark and baryonic) decreasing with time, possibly due to its gradual conversion into some other entity, generally referred to as dark energy. A generalized form of Brans-Dicke theory [27], where we have a variable BD parameter $\omega(\varphi)$ which is regarded as a function of scalar field parameter ($\varphi \equiv 1/G$), has been the basis of all calculations in the present model. One easily obtains the time dependence of gravitational constant from the time dependence of $\varphi$. An important finding of this study is that the gravitational constant gradually increases with time, as evident from many other studies not based on Brans-Dicke theory [35]. On the basis of our results we have graphically depicted the time variation of a quantity $\rho a^3$ which can be regarded as a measure of the matter content of the universe.

## Theoretical Model

The field equations in the generalized Brans-Dicke theory, for a spatially flat Robertson-Walker space-time, are given by [27],

$$3\left(\frac{\dot{a}}{a}\right)^2 = \frac{\rho}{\varphi} + \frac{\omega(\varphi)}{2}\left(\frac{\dot{\varphi}}{\varphi}\right)^2 - 3\frac{\dot{a}}{a}\frac{\dot{\varphi}}{\varphi}, \tag{1}$$

$$2\frac{\ddot{a}}{a} + \left(\frac{\dot{a}}{a}\right)^2 = -\frac{\omega(\varphi)}{2}\left(\frac{\dot{\varphi}}{\varphi}\right)^2 - 2\frac{\dot{a}}{a}\frac{\dot{\varphi}}{\varphi} - \frac{\ddot{\varphi}}{\varphi}. \tag{2}$$

Combining (1) and (2) one gets,

$$2\frac{\ddot{a}}{a}+4\left(\frac{\dot{a}}{a}\right)^2=\frac{\rho}{\varphi}-5\frac{\dot{a}}{a}\frac{\dot{\varphi}}{\varphi}-\frac{\ddot{\varphi}}{\varphi}. \quad (3)$$

Considering the conservation of matter of the universe we propose the following relation.

$$\rho=f(t)(\rho_0 a_0^3)a^{-3}=f(t)\rho_0 a^{-3}, \quad \text{(taking } a_0=1\text{)} \quad (4)$$

Here $a_0$ and $\rho_0$ are the scale factor and the matter density of the universe respectively at the present time. According to some studies, the matter content of the universe may not remain proportional to $\rho_0 a_0^3$ [31, 36]. There may be an inter-conversion between dark energy and matter (both baryonic and dark matter) [34, 36]. In the present model, a factor $f(t)$ has been introduced to account for the conversion of matter into dark energy or its reverse process. It is assumed here that this conversion, if there is any, is extremely slow. This assumption of slowness is based on the fact that there are studies where the variation of density of matter is expressed as $\rho=\rho_0 a^{-3}$, which actually indicates a conservation of the total matter content of the universe [27]. In the present calculations, the factor $f(t)$ is taken as a very slowly varying function of time, in comparison with the scale factor. Equation (4) makes it necessary that $f(t)=1$ at $t=t_0$ where $t_0$ denotes the present instant of time when the scale factor $a=a_0=1$ and the density $\rho=\rho_0$.

To make the differential equation (3) tractable, let us propose the following ansatz.

$$\varphi=\varphi_0 a^{-3} \quad (5)$$

Here $\varphi$ has been so chosen that it has the same dependence upon scale factor as that of the matter density. This choice of $\varphi$ makes the first term on the right hand side of equation (3) independent of the scale factor ($a$).

In equation (5) we have taken $\varphi=\varphi_0$ for $a=a_0=1$ .

Combining (3) and (5) and treating $f$ as a constant we have,

$$\frac{\ddot{a}}{a}-\left(\frac{\dot{a}}{a}\right)^2=-f\frac{\rho_0}{\varphi_0} \quad (6)$$

In terms of Hubble parameter $H=\frac{\dot{a}}{a}$, equation (6) takes the following form.

$$\left(\dot{H}+H^2\right)-H^2=\frac{dH}{dt}=-f\frac{\rho_0}{\varphi_0} \quad (7)$$

Integrating equation (7) and taking $H=H_0$ at $a=a_0=1$ we have,

$$H=\frac{\dot{a}}{a}=f\frac{\rho_0}{\varphi_0}(t_0-t)+H_0 \quad (8)$$

Integrating (8) and requiring that $a=a_0=1$ at $t=t_0$,

$$a=Exp\left[-\frac{1}{2}f\frac{\rho_0}{\varphi_0}(t^2+{t_0}^2)+\left(f\frac{\rho_0}{\varphi_0}t_0+H_0\right)t-H_0t_0\right] \quad (9)$$

In deriving the equations (8) and (9), $f$ has been treated as a constant assuming its extremely slow time variation compared to the scale factor. The time dependence of $f$ is determined later in this study and incorporated in equation (9).

Figure 1 shows the variation of scale factor $(a)$ and $f(t)$ as functions of scaled time $(t/t_0)$ where $t_0(= 14\, billion\, years)$ is the age of the universe. These curves show that these two parameters increases and decreases with time respectively and there is a long period of time over which $f(t)$ does not change appreciably. Figure 2 shows the variation of $f$ as a function of scale factor$(a)$ and this curve is consistent with our initial assumption about $f$, according to which it changes much less rapidly compared to the scale factor. It appears from the curve that our assumption remains valid nearly upto the present epoch where $= a_0 = 1$ , for the functional form chosen empirically for $f(t)$.

Using (9), the deceleration parameter $q\left(\equiv -\frac{\ddot{a}a}{\dot{a}^2}\right)$ becomes

$$q = -1 + \frac{f\rho_0/\varphi_0}{\left(\frac{f\rho_0}{\varphi_0}(t_0-t)+H_0\right)^2} . \tag{10}$$

Now letting $q = q_0$ at $t = t_0$ in (10), one obtains $q_0 = -1 + \frac{\rho_0}{H_0^2\varphi_0} = -0.9652$ .

Its negative sign shows that the universe is presently passing through a state of accelerated expansion and this fact is consistent with other studies.

Equation (10) clearly shows that a signature flip in $q$ takes place at $t = \tau$ where,

$$\tau = t_0 - \left(\sqrt{\frac{\varphi_0}{f\rho_0}} - H_0\frac{\varphi_0}{f\rho_0}\right) \tag{11}$$

Taking $\tau = \alpha t_0$ with $\alpha < 1$ we get the following quadratic equation from equation (11).

$$H_0x^2 - x + t_0(1-\alpha) = 0 \quad \text{with} \quad x = \sqrt{\frac{\varphi_0}{f\rho_0}} \tag{12}$$

To get a single value for $x$ we must have $\alpha = 1 - \frac{1}{4H_0t_0} = 0.757$ (13)

Thus we get, $x \equiv \sqrt{\frac{\varphi_0}{f\rho_0}} = \frac{1}{2H_0}$ leading to $f = \frac{4H_0^2\varphi_0}{\rho_0}$ (14)

The values of different cosmological parameters used in the present study are,

$H_0 = 72\left(\frac{Km}{Sec}\right) per\ Mega\ Parsec = 2.33\times10^{-18}sec^{-1}$, $t_0 = 14\, billion\, years = 4.415\times10^{17}sec,$ $\varphi_0 = \frac{1}{G} = 1.498\times10^{10}m^{-3}Kgs^2$, $\rho_0 = 2.831\times10^{-27}Kg/m^3$ (present density of dark matter + baryonic matter).

Let us now formulate the factor $f(t)$ from different criteria to be satisfied by it.

Based on the equations (13, 14), we may write,

$$f(t) = 4\frac{H_0^2\varphi_0}{\rho_0} \quad \text{at } at\ t = \alpha t_0 = t_0 - \frac{1}{4H_0} \tag{15}$$

According to an initial requirement (to satisfy equation 4), $f(t) = 1\ at\ t = t_0$ (16)

Let us now propose a relation between $f$ and $t$ which will satisfy the conditions expressed by (15) and (16). We may choose an empirical form, such as $f = Ae^{\beta t}$ which will not be negative if $A$ >0. Here we need to determine the values of the constants $A$ and $\beta$ from the conditions expressed by the equations (15) and (16). The values of these constants are thus found to be $A = (f_2)^{\frac{1}{1-\alpha}}$ and $\beta = -\frac{lnf_2}{1-\alpha}\frac{1}{t_0}$. Hence we have,

$$f = (f_2)^{\frac{1}{1-\alpha}} Exp\left[-\frac{lnf_2}{1-\alpha}\frac{t}{t_0}\right] \quad \text{with} \quad f_2 = 4\frac{H_0^2\varphi_0}{\rho_0} \tag{17}$$

This functional form of $f(t)$ keeps it positive which is a requirement of equation (4), since the density of matter can not be negative. This time dependent form of $f(t)$ has been used in all expressions in the present study.

Figure 3 shows the variation of the density of matter (both dark and baryonic) of the universe as a function of scale factor. It decreases more rapidly beyond a certain value of scale factor (around $a = 0.7$) where the signature flip of deceleration parameter takes place, as evident from Figure 9. It may be an indication for a greater rate of conversion of matter into dark energy beyond the signature flip of the deceleration parameter.

Figure 4 shows the variation of the Hubble parameter as a function of the scale factor.

According to Brans-Dicke theory, the gravitational constant is the reciprocal of the scalar field parameter $\varphi$. Therefore, using equations (5) and (9) we have,

$$G = \frac{1}{\varphi} = \frac{a^3}{\varphi_0} = \frac{1}{\varphi_0} Exp\left[-\frac{3}{2}\frac{f\rho_0}{\varphi_0}(t^2 + {t_0}^2) + 3H_0(t - t_0) + 3\frac{f\rho_0 t_0}{\varphi_0}t\right] \tag{18}$$

and the fractional change of $G$ per unit time is given by,

$$\frac{\dot{G}}{G} = \frac{1}{G}\ \frac{d}{dt}\left[\frac{1}{\varphi_0} Exp\left\{-\frac{3}{2}\frac{f\rho_0}{\varphi_0}(t^2 + {t_0}^2) + 3H_0(t - t_0) + 3\frac{f\rho_0 t_0}{\varphi_0}t\right\}\right] \tag{19}$$

According to Brans-Dicke theory, $G = \frac{1}{\varphi}$. Using this relation and equation (5) we get,

$$\left(\frac{\dot{G}}{G}\right)_{t=t_0} = 3H_0 = 2.2 \times 10^{-10}\ Yr^{-1} \tag{20}$$

According to a study by Weinberg [33], $\left(\frac{\dot{G}}{G}\right)_{t=t_0} \leq 4 \times 10^{-10}\ Yr^{-1}$ . Our result is consistent with this observation.

In the figures (5) and (6), we have plotted $G$ and $\frac{\dot{G}}{G}$ respectively as functions of the scale factor. The gravitational constant is found to increase with time with a varying rate. This increasing nature of $G$, with time, has been found in many other studies [28, 29, 30, 32, 35].

At $t = t_0$, $\frac{\dot{G}}{G}$ is positive, implying that the gravitational constant is presently increasing with time.

Using (2) and (5) we get,

$$\omega(\varphi) = -\frac{2}{3}\left(1 + \frac{\ddot{\varphi}\varphi}{\dot{\varphi}^2}\right) = -\frac{2}{9}\left(7 - \frac{\ddot{a}a}{\dot{a}^2}\right) = -\frac{2}{9}(7 + q). \qquad (21)$$

Equation (21) shows that the Brans-Dicke parameter $\omega(\varphi)$ has a linear relationship with the deceleration parameter $(q)$.

At $t = t_0$ we have,

$$\omega(\varphi_0) = -\frac{2}{9}(7 + q_0) = -1.341. \qquad (22)$$

Substituting for $q$ in equation (21) from equation (10)

$$\omega(\varphi) = -\frac{2}{9}(7 + q) = -\frac{2}{9}\left[6 + \frac{f\rho_0/\varphi_0}{\left[\frac{f\rho_0}{\varphi_0}(t_0 - t) + H_0\right]^2}\right]. \qquad (23)$$

Equation (23) shows the time variation of Brans-Dicke parameter $\omega(\varphi)$.

Combining the equations (5) and (9) one gets,

$$\varphi = \varphi_0 a^{-3} = \varphi_0 Exp\left[-3\left(-\frac{1}{2}\frac{f\rho_0 t^2}{\varphi_0} + \left(H_0 + \frac{f\rho_0 t_0}{\varphi_0}\right)t - H_0 t_0 - \frac{1}{2}\frac{f\rho_0 {t_0}^2}{\varphi_0}\right)\right]. \qquad (24)$$

Figures 7 shows the variation of $\varphi$ and $\omega$ as functions of scale factor $(a)$. Figure 8 shows the variation of Brans-Dicke parameter $\omega(\varphi)$ as a function of the scalar field $\varphi$. It is found to be negative over the entire range of study. It is evident from these figures that the most negative value of $\omega$ corresponds to the time of signature flip of deceleration parameter.

Figures 9 and 10 show respectively the variation of the deceleration parameter as functions of the scale factor and scaled time $(t/t_0)$. These curves show that there was a phase of decelerated expansion (with $q > 0$) of the universe which was preceded and followed by phases of acceleration.

Figures 11 and 12 depict respectively, with respect to scale factor $(a)$ and scaled time $(t/t_0)$, the variation of a quantity $\rho a^3$ which can be regarded as a measure of the matter content of the universe (for both dark and baryonic matter). These curves show that the matter content decreases due to its conversion into some other form of matter or energy responsible for the accelerated expansion of the universe.

## Conclusions

It is important to note from the findings of the present model that a generalized scalar tensor theory, where the BD parameter $\omega$ is regarded as a function of the scalar field $\varphi$, can account for a smooth transition of the universe from a phase of decelerated expansion to a phase of accelerated expansion, simply by taking into account an inter-conversion between matter (both baryonic and dark) and dark energy. In order to account for this phenomenon of inter-conversion, we have introduced a function $f(t)$ in this model with an assumption that the scale factor changes much more rapidly with time in comparison to this factor. Validity of this assumption is evident from figure 2. Its functional form has been determined empirically by using the information regarding its values at two different instants of time. To solve the field equations conveniently in an analytical way, we have assumed an empirical dependence of the BD scalar field parameter $\varphi$ on the scale factor $(a)$. The findings of this study show that the process of expansion started with acceleration which was followed by a phase of deceleration and again it has made a transition to its present state of acceleration, as evident from the present negative value of the deceleration parameter$(q)$, and it will continue to remain in the state of acceleration. The present model shows that the dark energy, which is regarded as responsible for the apparently strange accelerated expansion, is being produced at the cost of the matter of the universe (of both dark and baryonic form). As a consequence of this process the matter content of the universe decreases with time and it has been depicted graphically. This study also reveals the gravitational constant, reciprocal of the scalar field, increases with time. Its rate of fractional change per second is consistent with other studies in this regard. The present study shows the variation of the BD parameter $\omega(\varphi)$ graphically as a function of time and also the scalar field parameter $\varphi$. The variation of scale factor as a function of time and the variation of other parameters as functions of scale factor have been shown graphically. It has been shown graphically that there has been a decrease of matter content with time, indicating a conversion of matter into other forms of energy responsible for the accelerated expansion of the universe. By changing the assumption regarding the dependence of the scalar field $\varphi$ upon the scale factor $(a)$, one is likely to achieve an improvement over this model. A more rigorous study can be carried out in future with a different assumption regarding the function$f(t)$, where its functional form would be determined by incorporating an ansatz at the beginning of calculations regarding the dependence of $f(t)$ upon the scale factor $(a)$.

# FIGURES

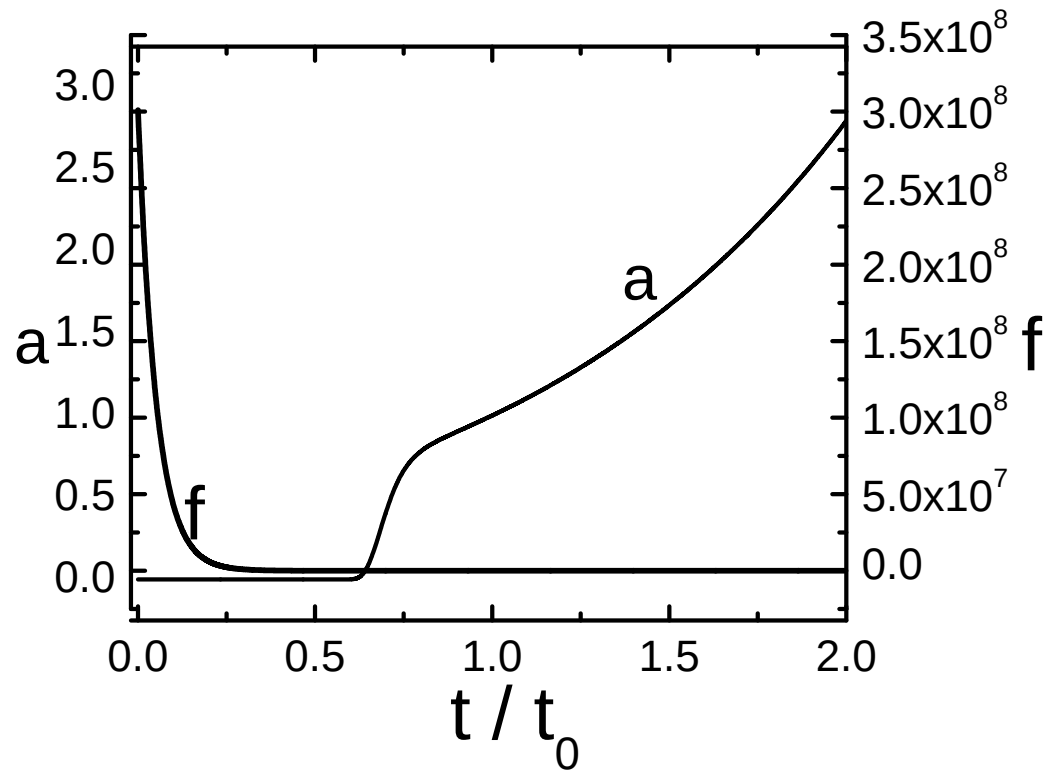

Figure 1. Variation of scale factor $(a)$ and $f(t)$ as functions of scaled time.

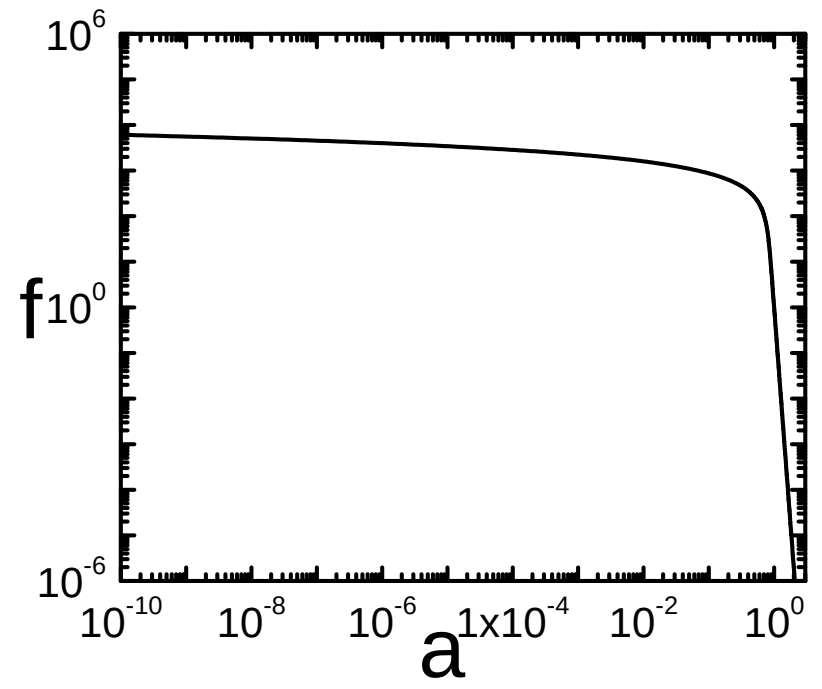

Figure 2. Variation of $f(t)$ as a function of scale factor $(a)$.

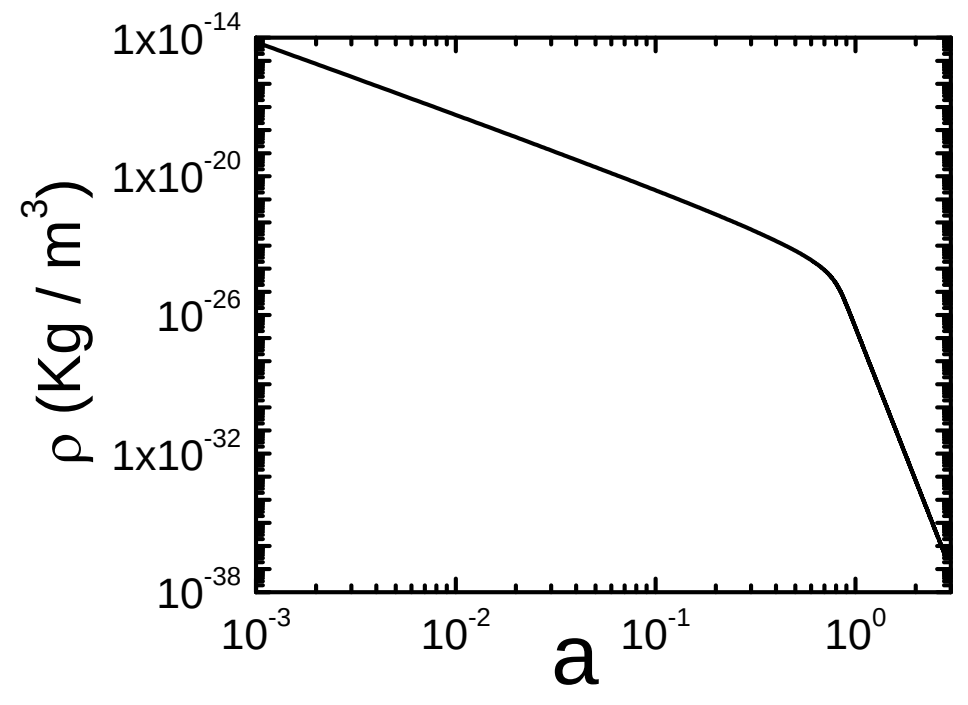

Figure 3. Variation of matter density $(\rho)$as a function of scale factor $(a)$.
$\rho_0 = 2.831 \times 10^{-27} Kg/m^3$

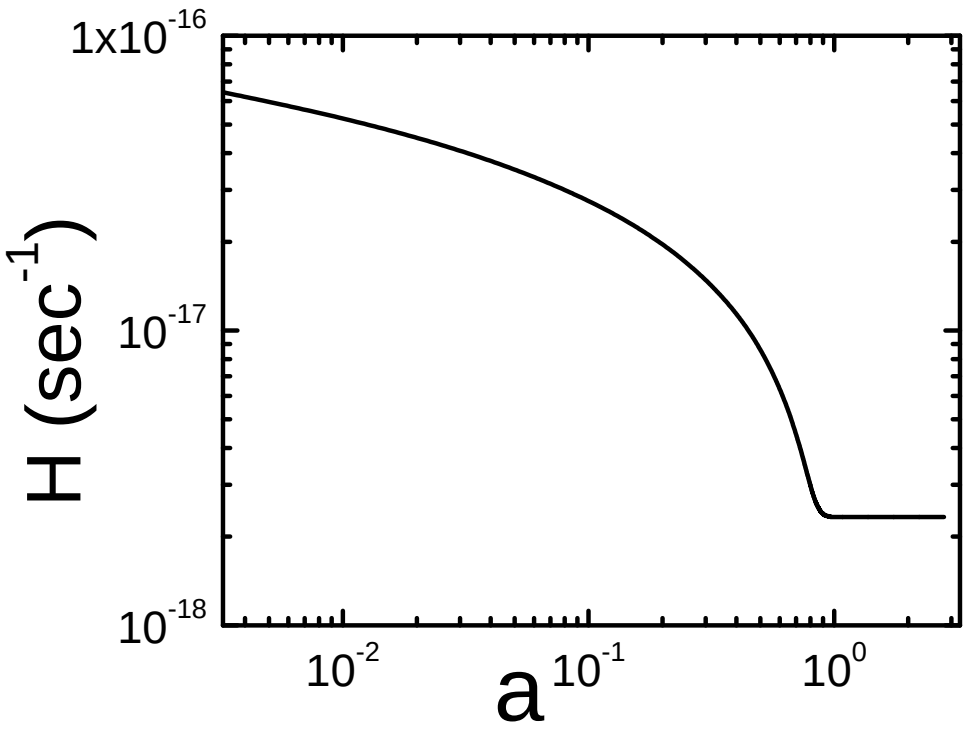

Figure 4. Variation of Hubble parameter as a function of scale factor $(a)$.

$$H_0 = 2.33 \times 10^{-18} sec^{-1}$$

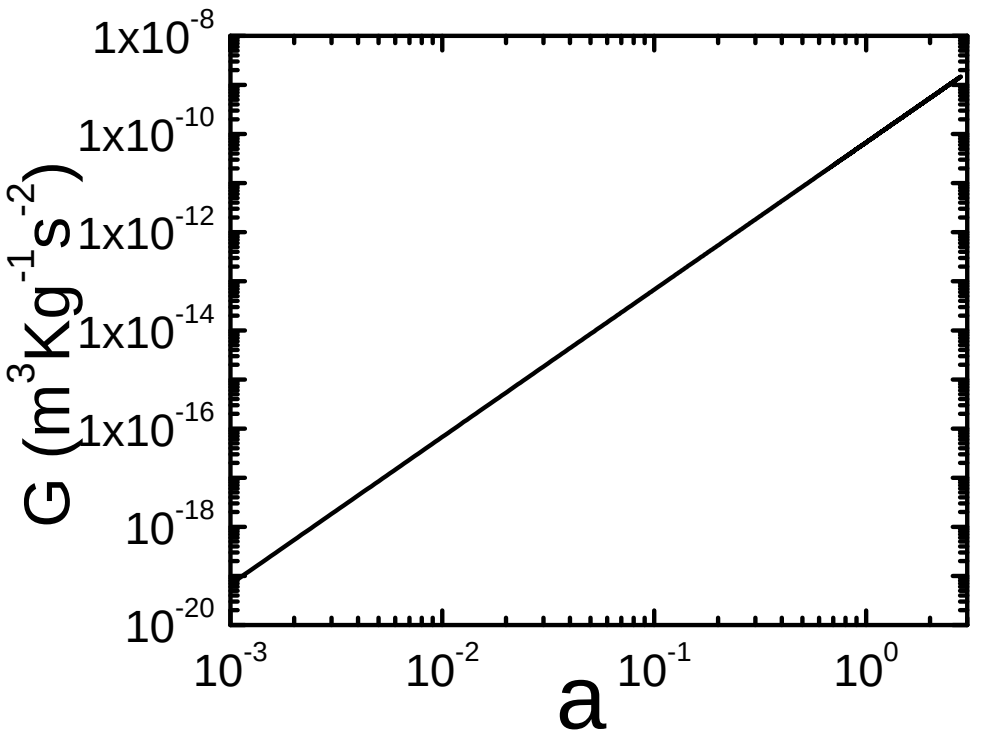

Figure 5. Variation of gravitational constant as a function of scale factor.

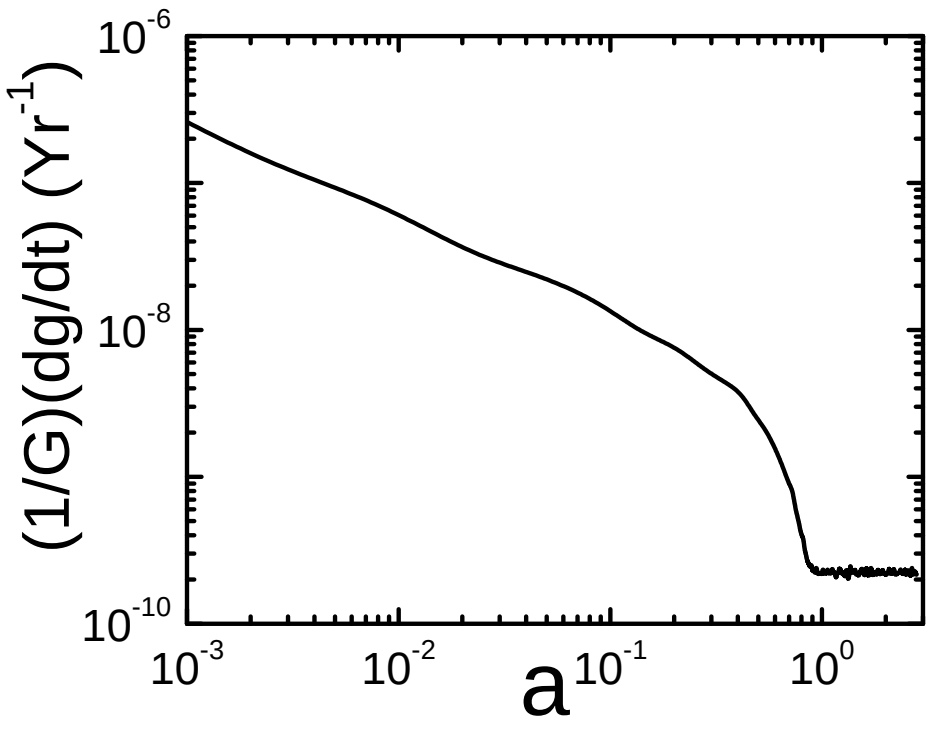

Figure 6. Variation of fractional change of $G$ per year, as a function of scale factor.

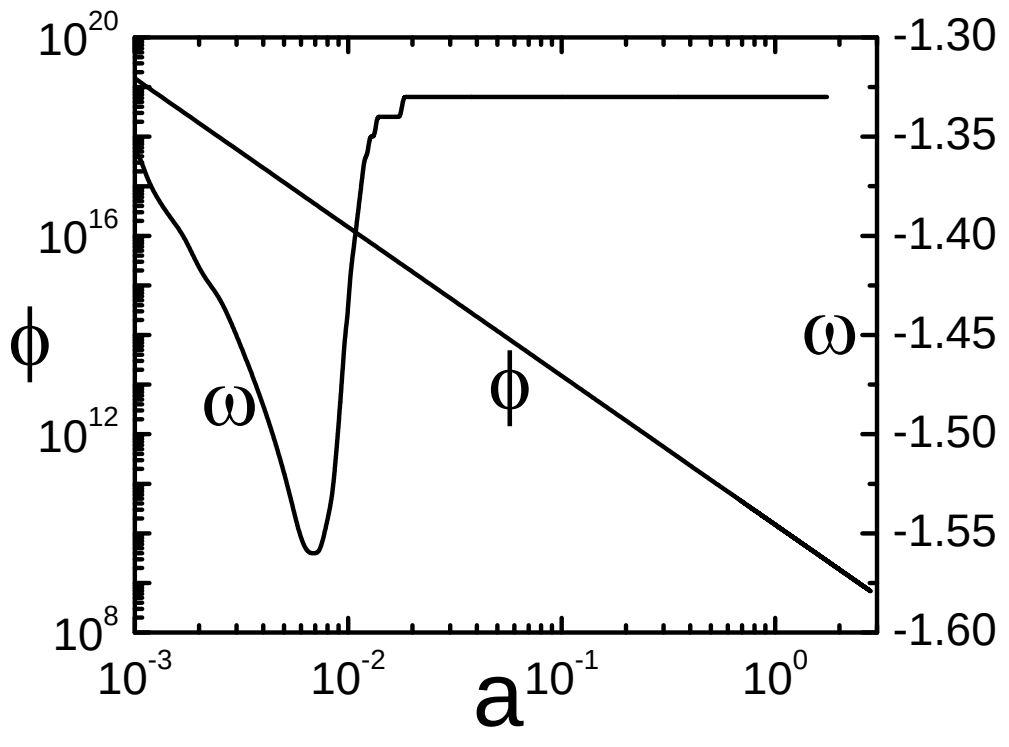

Figure 7. Variation of $\varphi$ and $\omega$ as functions of scale factor $(a)$.

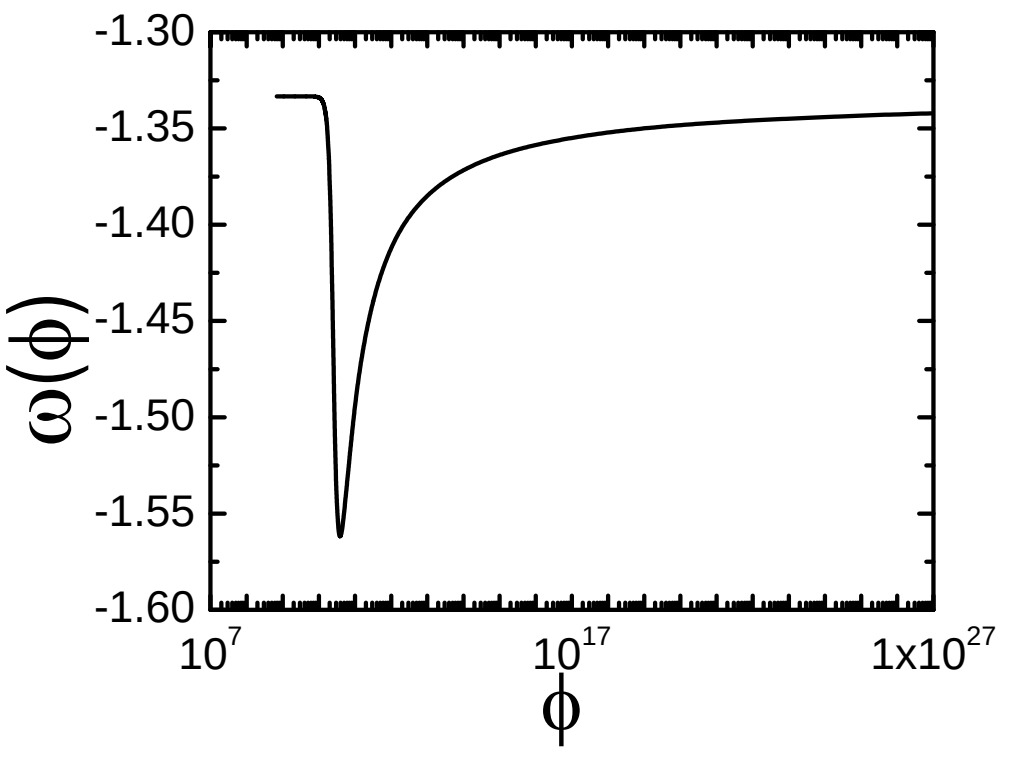

Figure 8. Variation of $\omega(\varphi)$ as a function of the scalar field $\varphi$.

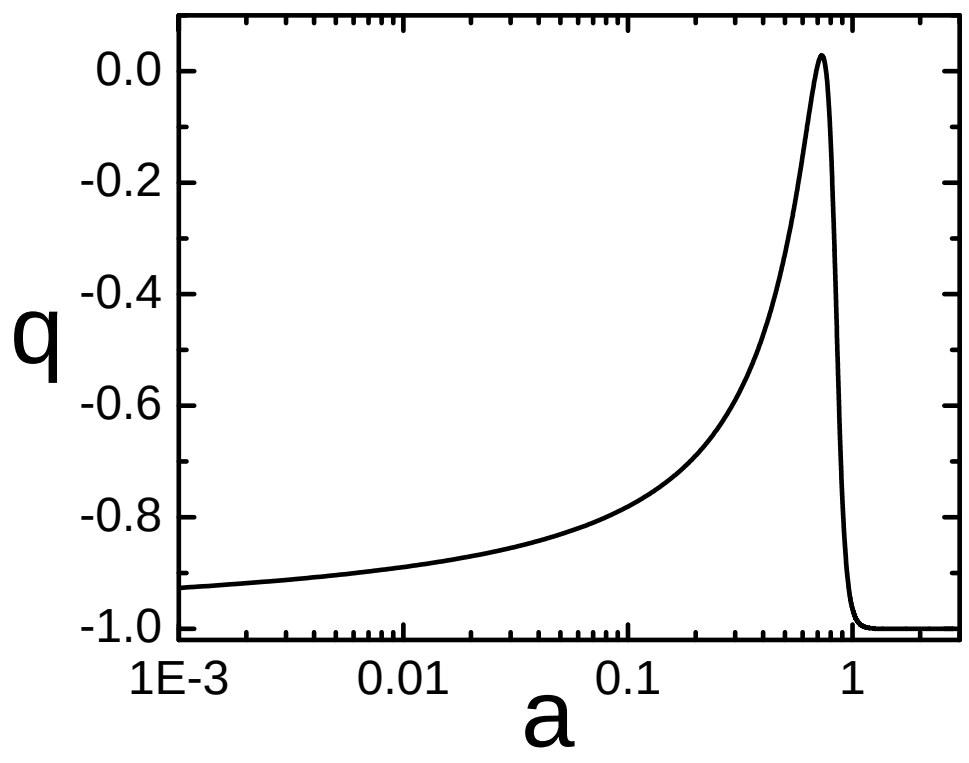

Figure 9. Variation of deceleration parameter as a function of scale factor.

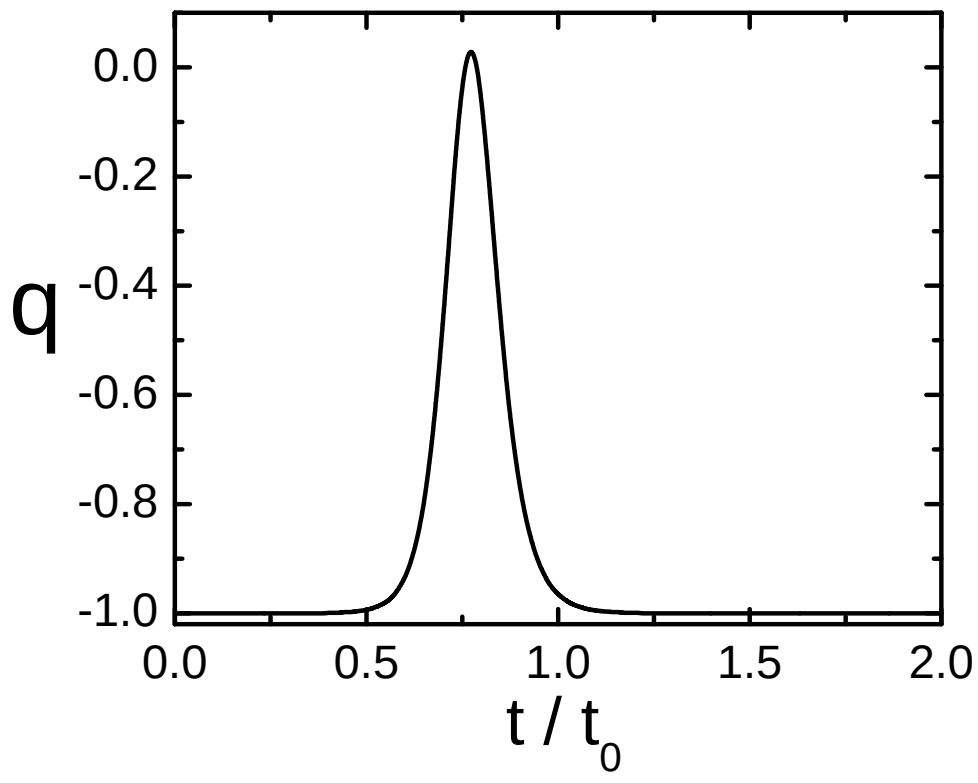

Figure 10. Variation of deceleration parameter as a function of scaled time.

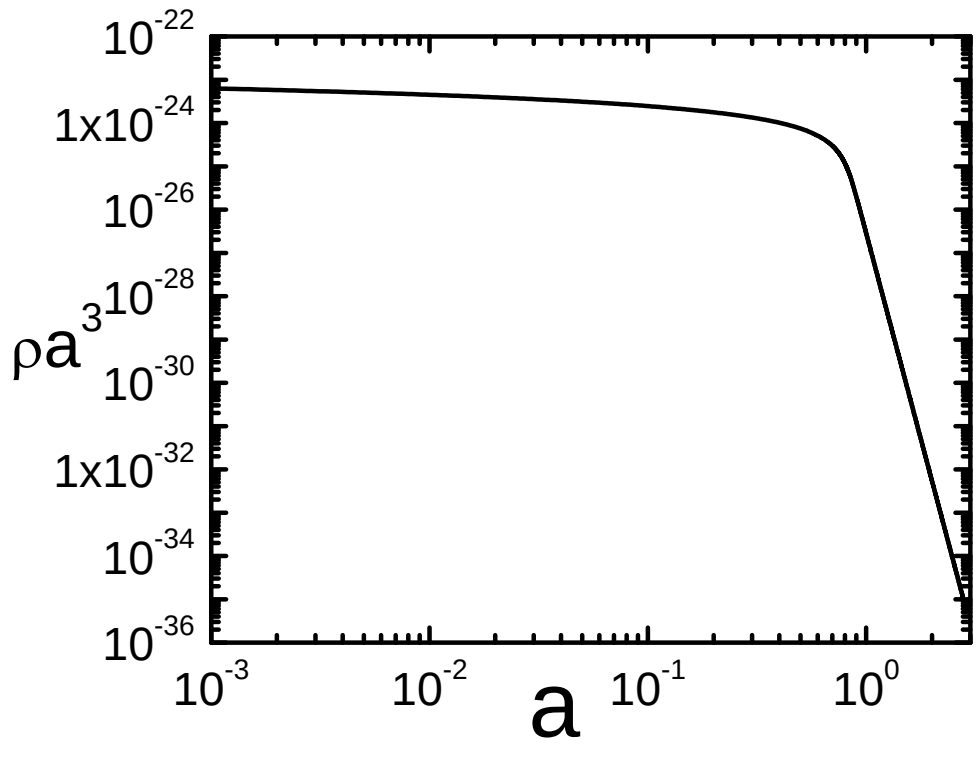

Figure 11. Variation of $\rho a^3$ as a function of scale factor ($a$).

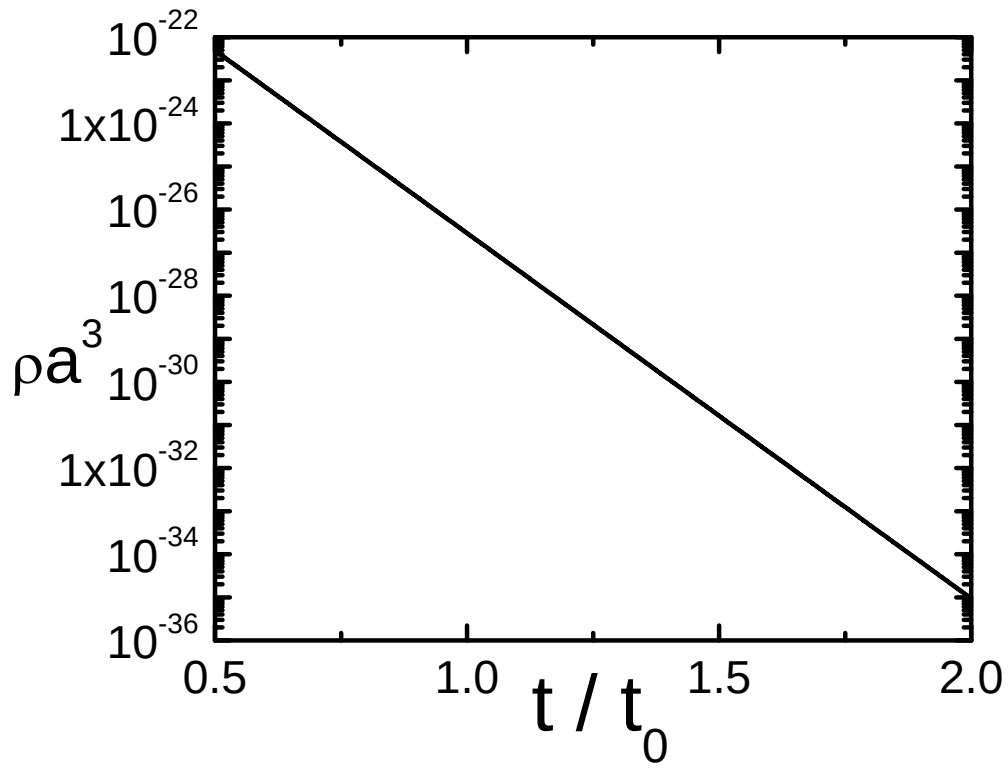

Figure 12. Variation of $\rho a^3$ as a function of scaled time.